\documentclass [a4paper,fleqn, 12pt]{article}
\usepackage{graphicx}
\usepackage[small]{subfigure,epsfig}

\usepackage {amsmath} \usepackage{amssymb} \usepackage{cite}

\newcommand{\pr}{\textbf{Proof.}\ }
\newcommand{\cn}

\begin{document}


\title
{Explicit expressions for meromorphic solution of autonomous nonlinear ordinary differential equations}

\author
{Maria V. Demina, \and Nikolay A. Kudryashov}

\date{Department of Applied Mathematics, National Research Nuclear University
MEPHI, 31 Kashirskoe Shosse,
115409 Moscow, Russian Federation}




\maketitle

\begin{abstract}

Meromorphic solutions of autonomous nonlinear ordinary differential equations are studied. An algorithm for constructing meromorphic solutions in explicit form is presented. General expressions for meromorphic solutions (including rational, periodic, elliptic) are found for a wide class of autonomous nonlinear ordinary differential equations.

\end{abstract}






\section{Introduction}

Autonomous nonlinear partial and ordinary differential equations
frequently arise in mathematical models describing processes and
phenomena in physics, chemistry, biology, etc. The problem of
constructing exact solutions to autonomous nonlinear differential
equations is intensively studied
\cite{Kudr88, Kudr90, Kudr91, Kudr92, Vernov01, Vernov02,
Hone01, Kudryashov05, Kudryashov06, Kudr05b}. A number of
methods and algorithms has been introduced. As a rule these methods
deal with a priori fixed families of meromorphic exact solutions.
Consequently, solutions outside given families may be missed. Very
often various methods yield the same families of exact solutions
written in a different way. And as a result many "new" exact
solutions appear. \cite{Kudr08a, Kudryashov01, Kudryashov02, Kudryashov03}
Thus an important problem is to classify families of exact
solutions. In this paper we study the general form of meromorphic
solutions to autonomous nonlinear ordinary differential equations.
The main aim is to present our algorithm, which can be applied to
construct exact meromorphic solutions in explicit form. For a wide
class of nonlinear ordinary differential equations we make a
conclusion that exact solutions given in our paper are the only
possible meromorphic solutions.

This paper is organized as follows. In section 2 we describe our
method and present explicit expressions for meromorphic solutions.
In sections 3 we give an example and construct meromorphic solutions
of a certain second order nonlinear differential equation.

\section{Method applied}

Let us look for nonconstant meromorphic exact solutions of an autonomous nonlinear ordinary differential
equation
\begin{equation}
\label{EQN} E[w(z)]=0.
\end{equation}
In this expression $E[w(z)]$ is a polynomial in $w(z)$ and its derivatives.
For every solution $w(z)$ of equation \eqref{EQN} there
exists a family of solutions $w(z-z_0)$. Without loss of generality,
we omit arbitrary constant $z_0$. Suppose the following.

\textit{Condition I.} For solutions of equation \eqref{EQN} there
are only $N$ different asymptotic expansions corresponding to the
Laurent series in a neighborhood of $z=0$
\begin{equation}
\begin{gathered}
\label{Laurent_expantion1}
w^{(i)}(z)=\sum_{k=1}^{p_i}\frac{c_{-k}^{(i)}}{z^k}+\sum_{k=0}^{\infty}c_k^{(i)}z^k,\quad
0<|z|<\varepsilon_i,\quad i=1, \ldots, N.
\end{gathered}
\end{equation}
In this expression $p_i>0$ is an order of the pole $z=0$. If $N=1$
we shall omit the upper index.

\textit{Condition II.} Substituting $w(z)=\lambda W(z)$ into equation \eqref{EQN} yields expression with only one term of the
highest degree in respect of~$\lambda$.

Note that if $w(z)$ has more than $N$ poles, then $w(z)$ is
periodic. Indeed, there exist poles $z=z_1$, $z=z_2$ of $w(z)$ such
that the functions $w(z+z_1)$, $w(z+z_2)$ are meromorphic solutions
of equation \eqref{EQN} with a pole at $z=0$. Thus we have the
equality $w(z)=w(z+z_2-z_1)$.

Meromorphic solutions of equation \eqref{EQN} are classified in the
following theorems.

\textbf{Theorem 1.} All meromorphic solutions of equation \eqref{EQN}
satisfying condition I with $N=1$ and condition II are of the form:
\newline 1) elliptic solutions with the periods $2\omega_1$, $2\omega_2$
\begin{equation}
\begin{gathered}
\label{Ex_Sol_Elliptic2} w(z)=\left\{\sum_{k=2}^{p}\frac{(-1)^k
c_{-k}}{(k-1)!}\frac{d^{k-2}}{dz^{k-2}}\right\}\wp(z;\omega_1,\omega_2)+
h_0,
\end{gathered}
\end{equation}
Necessary condition for elliptic solutions to exist is $c_{-1}=0$.
\newline 2) periodic solutions with the period $T$
\begin{equation}
\begin{gathered}
\label{Ex_Sol_Expp} w(z)=
\frac{\pi}{T}\left\{\sum_{k=1}^{p}\frac{(-1)^{k-1}
c_{-k}}{(k-1)!}\frac{d^{k-1}}{dz^{k-1}}\right\}\cot \left(\frac{\pi
z}{T}\right)+ h_0.
\end{gathered}
\end{equation}
\newline 3) rational solutions
\begin{equation}
\begin{gathered}
\label{Rat_sol_N=1}
w(z)=\sum_{k=1}^{p}\frac{c_{-k}}{z^k}+\sum_{k=0}^{m}c_kz^k,\quad
m\geq 0.
\end{gathered}
\end{equation}
In 1) and 2) $h_0$ is a constant.

\pr
For any meromorphic function $f(z)$ with the poles $\{a_n\}$
Mittag--Leffler found the representation \cite{Lavrentiev01}
\begin{equation}
\begin{gathered}
\label{Meromorphic_function}
f(z)=\sum_{n=1}^{\infty}\left\{P_n\left(\frac{1}{z-a_n}\right)-g_n(z)\right\}+h(z),
\end{gathered}
\end{equation}
where $\{P_n[(z-a_n)^{-1}]\}$ are principal parts of Laurent
expansions for the function $f(z)$ around poles $\{a_n\}$,
$\{g_n(z)\}$ are polynomials, and $h(z)$ is an entire function. The
polynomial $g_n(z)$ is added to the $n^{\text{th}}$ term of the sum
\eqref{Meromorphic_function} in order to provide convergence of the
series. A series of the form \eqref{Meromorphic_function} is said to
be convergent in $M\subset \mathbb{C}$ if only finite number of
terms possesses poles in $M$ and without these terms the series
converges \cite{Lavrentiev01}. Suppose $w(z)$ has two or more poles.
Then from condition I follows that $w(z)$ is doubly periodic
(elliptic) or simply periodic. For the elliptic function $w(z)$ with
Laurent expansion \eqref{Laurent_expantion1} expression
\eqref{Meromorphic_function} reads
\begin{equation}
\begin{gathered}
\label{Periodic_Elliptic}
w(z)=\frac{c_{-2}}{z^2}+{\sum_{(n,\,m)\neq(0,\,0)}}
\left[\frac{c_{-2}}{(z-2n\omega_1-2m\omega_2)^2}-\frac{c_{-2}}{(n\omega_1+m\omega_2)^2}\right]\\
+\sum_{k=3}^{p}\sum_{n,\,m}^{}\frac{c_{-k}}{(z-2n\omega_1-2m\omega_2)^k}+
h_0.
\end{gathered}
\end{equation}
Solution \eqref{Periodic_Elliptic} can be rewritten in terms of
Weierstrass elliptic function $\wp$ satisfying the equation
\begin{equation}
\begin{gathered}
\label{Wier} (\wp_z)^2=4\wp^3-g_2\wp-g_3.
\end{gathered}
\end{equation}
Thus we obtain representation for the elliptic solution $w(z)$ in
the form \eqref{Ex_Sol_Elliptic2}. Necessary condition $c_{-1}=0$
follows from the theorem for total sum of the residues  of an
elliptic function in the parallelogram of periods. Writing
expression \eqref{Meromorphic_function} for the solution $w(z)$ with
period $T$ yields
\begin{equation}
\begin{gathered}
\label{Periodic_exp} w(z)=\frac{c_{-1}}{z}+\sum_{n\neq
0}^{}\left[\frac{c_{-1}}{(z-nT)}+\frac{c_{-1}}{nT}\right]+
\sum_{k=2}^{p}\sum_{n}^{}\frac{c_{-k}}{(z-nT)^k}+ h(z).
\end{gathered}
\end{equation}
In the paper \cite{Eremenko01} meromorphic traveling wave solutions
of the Kuramoto -- Sivashinsky equation were studied. A powerful
method based on Nevanlinna theory \cite{Nevanlinna01} was developed.
From the results of the paper \cite{Eremenko01} it follows that
periodic solutions of equation \eqref{EQN} satisfying condition II
are rational in $e^{rz}$. Making change of variables $s=e^{rz}$,
$y(s)=w(z)$ in equation \eqref{EQN} and then substituting $y=s^n$,
$n>0$ as $s$ tends to infinity and $y=s^{-l}$, $l>0$ as $s$ tends to
zero into the result, we wee that an entire periodic function $h(z)$
in \eqref{Periodic_exp} is a constant since condition II is valid.
Further, making use of the formula
\begin{equation}
\begin{gathered}
\label{Periodic_sols_cot} \frac{1}{z}+\sum_{n\neq
0}^{}\left[\frac{1}{(z-nT)}+\frac{1}{nT}\right]=\frac{\pi}{T}\cot
\left(\frac{\pi z}{T}\right)
\end{gathered}
\end{equation}
we get \eqref{Ex_Sol_Expp} from \eqref{Periodic_exp} with
$h(z)=h_0$. Now suppose $w(z)$ has only one pole, then
\begin{equation}
\begin{gathered}
\label{Rational_proof} w(z)=\sum_{k=1}^{p}\frac{c_{-k}}{z^k}+h(z),
\end{gathered}
\end{equation}
where an entire function $h(z)$ does not have an essential
singularity at infinity \cite{Eremenko01}. This completes the proof.

\textit{Remark.} For fixed values of parameters in equation
\eqref{EQN}, if any, there may exist only one meromorphic solution
(rational, simply periodic or elliptic) with  a pole at $z=0$.

In the next theorem we generalize results of theorem 1 to
the case $N>1$. Let us call poles of meromorphic solution $w(z)$
with the Laurent expansion $w^{(i)}(z)$ (see
\eqref{Laurent_expantion1}) as poles of type $i$.

\textbf{Theorem 2.} All meromorphic solutions of equation \eqref{EQN}
satisfying condition I and condition II are of the form:
\newline 1) elliptic solutions with the periods $2\omega_1$, $2\omega_2$
\begin{equation}
\begin{gathered}
\label{Ex_Sol_EllipticN} w(z)=\left\{\sum_{i\in\,
I}^{}\sum_{k=2}^{p_i}\frac{(-1)^k
c_{-k}^{(i)}}{(k-1)!}\frac{d^{k-2}}{dz^{k-2}}\right\}\left(\frac14\left[
\frac{\wp_z(z)+B_i}{\wp(z)-A_i}\right]^2-\wp(z)\right)\\
+\sum_{i\in \,
I}^{}\frac{c_{-1}^{(i)}(\wp_z(z)+B_i)}{2\,(\wp(z)-A_i)}+\left\{\sum_{k=2}^{p_{i_0}}\frac{(-1)^k
c_{-k}^{(i_0)}}{(k-1)!}\frac{d^{k-2}}{dz^{k-2}}\right\}\wp(z)+ h_0,
\end{gathered}
\end{equation}
where $\wp(z)\stackrel{def}{=}\wp(z;\omega_1,\omega_2)$,
$B_i^2=4A_i^3-g_2A_i-g_3$. Necessary condition for elliptic
solutions \eqref{Ex_Sol_EllipticN} to exist is
\begin{equation}\label{Condition_ellipticN}
\sum_{i\in\,I}^{}c_{-1}^{(i)}+c_{-1}^{(i_0)}=0.
\end{equation}
\newline 2) periodic solutions with the period $T$
\begin{equation}
\begin{gathered}
\label{Ex_Sol_ExppN} w(z)=\frac{\pi}{T}\left\{\sum_{i\in\,
I}^{}\sum_{k=1}^{p_i}\frac{(-1)^{k-1}
c_{-k}^{(i)}}{(k-1)!}\frac{d^{k-1}}{dz^{k-1}}\right\}\frac{A_i\cot
\left(\frac{\pi z}{T}\right)+\frac{\pi}{T}}{A_i-\frac{\pi}{T}\cot
\left(\frac{\pi
z}{T}\right)}\\
+\frac{\pi}{T}\left\{\sum_{k=1}^{p_{i_0}}\frac{(-1)^{k-1}
c_{-k}^{(i_0)}}{(k-1)!}\frac{d^{k-1}}{dz^{k-1}}\right\}\cot
\left(\frac{\pi z}{T}\right)+ h_0.
\end{gathered}
\end{equation}
\newline 3) rational solutions
\begin{equation}
\begin{gathered}
\label{Rat_sol_N}
w(z)=\sum_{k=1}^{p_{i_0}}\frac{c_{-k}^{(i_0)}}{z^k} +\sum_{i\in\,
I}^{}\sum_{k=1}^{p_i}\frac{c_{-k}^{(i)}}{(z-a_i)^k}+\sum_{k=0}^{m}h_kz^k,\quad
m\geq 0.
\end{gathered}
\end{equation}
In 1), 2) and 3) $\{A_i\}$, $\{B_i\}$, $\{a_i\}$, $\{h_i\}$ are
constants and $I=\varnothing$ or $I$ $\subseteq$ $\{1,2,\ldots, N\}$
$\setminus\{i_0\}$, $1\leq i_0 \leq N$.

\pr Let $w(z)$ be a meromorphic solution of equation \eqref{EQN}
with poles of types $j$, $j\in \, J$, $J\subseteq$ $\{1,2,\ldots,
N\}$. Without loss of generality, we may select any $i_0\in$ $J$ and
suppose that the point $z=0$ is a pole of type $i_0$ for $w(z)$. Let
us set  $I=J\setminus\{i_0\}$. For $N=1$ we see that
$I=\varnothing$. Let us consider the case $N>1$. If there are more
than $|I|+1$ poles of different types, then $w(z)$ is periodic
(simply periodic or elliptic). By analogy with the case $N=1$ we get
the expression for elliptic solutions
\begin{equation}
\begin{gathered}
\label{Ex_Sol_EllipticN_proof} w(z)=\sum_{i\in
I\cup\{i_0\}}^{}c_{-1}^{(i)}\zeta(z-a_i)+\left\{\sum_{i\in
I\cup\{i_0\}}^{} \sum_{k=2}^{p_i}\frac{(-1)^k
c_{-k}^{(i)}}{(k-1)!}\frac{d^{k-2}}{dz^{k-2}}\right\}\wp(z-a_i)+
\tilde{h}_0,
\end{gathered}
\end{equation}
where $\zeta(z)$ is the Weierstrass $\zeta$--function and $a_{i_0}$
may be taken as zero. Note that $a_i\neq a_j$, $i$, $j$  $\in I$
$\cup$ $\{i_0\}$, $i\neq j$ are poles of elliptic solutions
\eqref{Ex_Sol_EllipticN_proof} in the parallelogram of periods built
on $2\omega_1$, $2\omega_2$. The function $\zeta(z)$ has a simple
pole at the point $z=0$ with residue $1$. The theorem for total sum
of the residues of an elliptic function in the parallelogram of
periods yields condition \eqref{Condition_ellipticN}. Using the
addition formulae \cite{Sikorsky01}
\begin{equation}
\begin{gathered}
\label{Addition_formulae} \zeta(z-a_i)=\zeta(z)-\zeta(a_i)+\frac12
\frac{\wp_z(z)+\wp_z(a_i)}{\wp(z)-\wp_(a_i)},\hfill \\
\wp(z-a_i)=-\wp(z)-\wp(a_i)+\frac14
\left[\frac{\wp_z(z)+\wp_z(a_i)}{\wp(z)-\wp_(a_i)}\right]^2
\end{gathered}
\end{equation}
and notation $A_i\stackrel{def}{=}\wp(a_i)$,
$B_i\stackrel{def}{=}\wp_z(a_i)$, we get equality
\eqref{Ex_Sol_EllipticN}. Further, from equation \eqref{Wier} we
have $B_i^2=4A_i^3-g_2A_i-g_3$. Note that for the parameters $h_0$
and $\tilde{h}_0$ the following correlation holds
\begin{equation}
\begin{gathered}
\label{Proof_ho}h_0=\tilde{h}_0- \sum_{i\in\, I}^{}c_{-1}^{(i)}
\zeta(a_i)-\sum_{i\in \,I}^{}c_{-2}^{(i)} \wp(a_i).
\end{gathered}
\end{equation}
Similarly to the case $N=1$ simply periodic solutions can be written
as
\begin{equation}
\begin{gathered}
\label{Ex_Sol_ExppN_proof} w(z)= \frac{\pi}{T}\left\{\sum_{i\in
I\cup\{i_0\}}^{}\sum_{k=1}^{p_i}\frac{(-1)^{k-1}
c_{-k}^{(i)}}{(k-1)!}\frac{d^{k-1}}{dz^{k-1}}\right\}\cot
\left(\frac{\pi (z-a_i}{T}\right)+ h_0,
\end{gathered}
\end{equation}
where again we may set $a_{i_0}=0$ and $a_i\neq a_j$, $i$, $j$ $\in
I$ $\cup$ $\{i_0\}$, $i\neq j$ are poles of simply periodic
solutions \eqref{Ex_Sol_EllipticN_proof} in a stripe built on $T$.
With the help of addition formulae for $\cot[\pi (z-a_i)T^{-1}]$ we
obtain expression \eqref{Ex_Sol_ExppN}, where
$A_i\stackrel{def}{=}\pi\cot[\pi a_iT^{-1}]T^{-1}$. For rational
solutions we have the equality
\begin{equation}
\begin{gathered}
\label{Rat_sol_N_proof} w(z)=\sum_{i\in
I\cup\{i_0\}}^{}\sum_{k=1}^{p_i}\frac{c_{-k}^{(i)}}{(z-a_i)^k}+\sum_{k=0}^{m}h_kz^k,\quad
m\geq 0,
\end{gathered}
\end{equation}
where $a_i\neq a_j$, $i$, $j$ $\in I$ $\cup$ $\{i_0\}$, $i\neq j$.
Without loss of generality, we set $a_{i_0}=0$ in
\eqref{Rat_sol_N_proof} to obtain \eqref{Rat_sol_N}.

\textit{Remark 1.} Again for fixed values of parameters in equation
\eqref{EQN}, if any, there may exist only one meromorphic solution
(rational, simply periodic or elliptic) with a pole at $z=0$ of type
$i$, $1\leq i \leq N$.

\textit{Remark 2.} While making classification of meromorphic
solutions, we should consider all the variants for $J\subseteq$
$\{1,2,\ldots, N\}$.

Now let us discuss the problem of finding exact meromorphic
solutions in explicit form. The main idea is to expand exact
solutions given in theorems 1 and 2 around their
poles and to compare coefficients of these expansions with
coefficients of series \eqref{Laurent_expantion1}. Our algorithm is
the following.

At \textit{the first step} one constructs asymptotic expansions
corresponding to Laurent series in a neighborhood of a pole $z=0$.

At \textit{the second step} one selects an expression for a
meromorphic solution (see theorem 1).

At \textit{the third step} one expands the meromorphic solution in a
neighborhood of poles $z=0$, $z=a_i$, $i\in I$.

At \textit{the fourth step} one compares coefficients of the series
found at the first and the third steps and solves an algebraic
system for the parameters of exact meromorphic solution. Along with
this, correlations for the parameters of equation \eqref{EQN} may
arise. If the meromorphic solution possesses poles of different
types, it is convenient to consider all possible Laurent expansions.
Optimal number of equations in algebraic system equals to the number
of parameters of the meromorphic solution and equation \eqref{EQN}
plus $1$. If this system is inconsistent, then equation \eqref{EQN}
does not possess meromorphic solutions with supposed expression.

At \textit{the fifth step} one verifies the meromorphic solution
obtained at the previous step, substituting this solution and
correlations on the parameters of equation \eqref{EQN} into the
latter.

In order to find the Laurent expansions for elliptic solutions
(theorem 1) in a neighborhood of poles $z=0$, $z=a_i$,
$i\in I$ one takes formula \eqref{Ex_Sol_EllipticN_proof} with
$a_{i_0}=0$ and then introduces notation
$A_i\stackrel{def}{=}\wp(a_i)$, $B_i\stackrel{def}{=}\wp_z(a_i)$,
$i\in I$. The parameters of elliptic solutions
\eqref{Ex_Sol_EllipticN} that one needs to calculate are $g_2$,
$g_3$, $h_0$, $A_i$, $i\in I$.

As far as simply periodic solutions (theorem 1) are
concerned, one takes formula \eqref{Ex_Sol_ExppN_proof} with
$a_{i_0}=0$, constructs Laurent expansions in a neighborhood of
poles $z=0$, $z=a_i$, $i\in I$, and introduces notation
$L\stackrel{def}{=}\pi^2 T^{-2}$, $A_i\stackrel{def}{=}\pi\cot[\pi
a_iT^{-1}]T^{-1}$, $i\in I$. The parameters of simply periodic
solutions \eqref{Ex_Sol_ExppN} to be found are the following $L$,
$h_0$, $A_i$, $i\in I$.

For rational solutions (theorem 2) the parameters to be
calculated are $h_0$, $a_i$, $i\in I$. In order to find the highest
exponent $m$ (see \eqref{Rat_sol_N}) one may construct the Laurent
expansion in a neighborhood of infinity.

To conclude this section we would like to emphasize that our method
is also applicable to algebraic autonomous nonlinear ordinary
differential equations with arbitrary coefficients in asymptotic
expansions corresponding to the Laurent series in a neighborhood of
poles. Arbitrary coefficients should be added to the list of
parameters. However, these equations may possess meromorphic
solutions of more complicated structure.

\section{Explicit construction of meromorphic solutions}

As an example let us find exact meromorphic solutions of the
following second order autonomous nonlinear differential equation
\begin{equation}
\begin{gathered}
\label{Meromorphic_sols_example_equation_first} w_{zz}+ww_z+\sigma
w^3+\delta w_z+\beta w^2+ \alpha w +\gamma=0,\quad \sigma\neq0.
\end{gathered}
\end{equation}
Without loss of generality, we may set $\delta=0$, $\sigma=-6$, and
study the equation
\begin{equation}
\begin{gathered}
\label{Meromorphic_sols_example_equation} w_{zz}+ww_z-6w^3+\beta
w^2+ \alpha w +\gamma=0.
\end{gathered}
\end{equation}
This equation admits two different asymptotic expansions
corresponding to Laurent series in a neighborhood of poles. They are
the following
\begin{equation}
\begin{gathered}
\label{Meromorphic_sols_example_Exp1}w^{(1)}(z)=\frac{1}{2z}+\frac{\beta}{20}+\left(\frac{11\beta^2}{200}+\alpha\right)\frac{z}{9}+
\left(\frac{\beta^3}{375}+\frac{\alpha\beta}{15}+\gamma\right)\frac{z^2}{2}+o(|z|^2)
\end{gathered}
\end{equation}
\begin{equation}
\begin{gathered}
\label{Meromorphic_sols_example_Exp2}w^{(2)}(z)=-\frac{2}{3z}+\frac{2\beta}{33}-
\left(\frac{20\beta^2}{363}+\alpha\right)\frac{z}{12}+\frac{3}{20}
\left(\frac{2\beta^3}{1331}+22\alpha\beta+\gamma\right)z^2+o(|z|^2)
\end{gathered}
\end{equation}
We see that equation \eqref{Meromorphic_sols_example_equation}
satisfies conditions II and I with $N=2$. First let us construct
simply periodic solutions possessing poles of single type (with
residue $c^{(1)}_{-1}=1/2$ and then with residue
$c^{(2)}_{-1}=-2/3$). In this case we should take exact meromorphic
solution in the form
\begin{equation}
\begin{gathered}
\label{Meromorphic_sols1} w(z)= c_{-1}\sqrt{L}\cot \left(
\sqrt{L}z\right)+ h_0,\quad L\stackrel{def}{=}\frac{\pi^2}{T^2}.
\end{gathered}
\end{equation}
Note that we omit arbitrary constant $z_0$. Finding the Laurent
expansion of function \eqref{Meromorphic_sols1} at the point $z=0$
\begin{equation}
\begin{gathered}
\label{Meromorphic_sols_example_Exp3}
w(z)=\frac{c_{-1}}{z}+h_0-\frac{c_{-1}L}{3}z-\frac{c_{-1}L^2}{45}z^3-
\frac{2c_{-1}L^3}{945}z^5+o\left(|z|^5\right)
\end{gathered}
\end{equation}
and comparing coefficients of this series with coefficients of
expansion \eqref{Meromorphic_sols_example_Exp1} (and then
\eqref{Meromorphic_sols_example_Exp2}), we obtain parameters $L$,
$h_0$ and correlations on the parameters $\alpha$, $\beta$,
$\gamma$. Substituting obtained meromorphic solutions into equation
\eqref{Meromorphic_sols_example_equation}, we make sure that this
equation indeed possesses meromorphic solutions of the form
\eqref{Meromorphic_sols1}. For the Laurent series
\eqref{Meromorphic_sols_example_Exp1} these solutions can be written
as
\begin{equation}
\begin{gathered}
\label{Meromorphic_sols_example_Sol1} w(z)=\frac{\pi}{2T}\cot
\left\{\frac{\pi(z-z_0)}{T}\right\}+\frac{\beta}{20}, \quad
T=10\left(-\frac{3}{11\beta^2+200\alpha}\right)^{1/2}\pi
\end{gathered}
\end{equation}
\begin{figure}[t]
 \centerline{
 \subfigure[$\alpha=-10$, $\beta=10$, $z_0=0$, $T=\sqrt{3}\pi/ 3$]{\epsfig{file=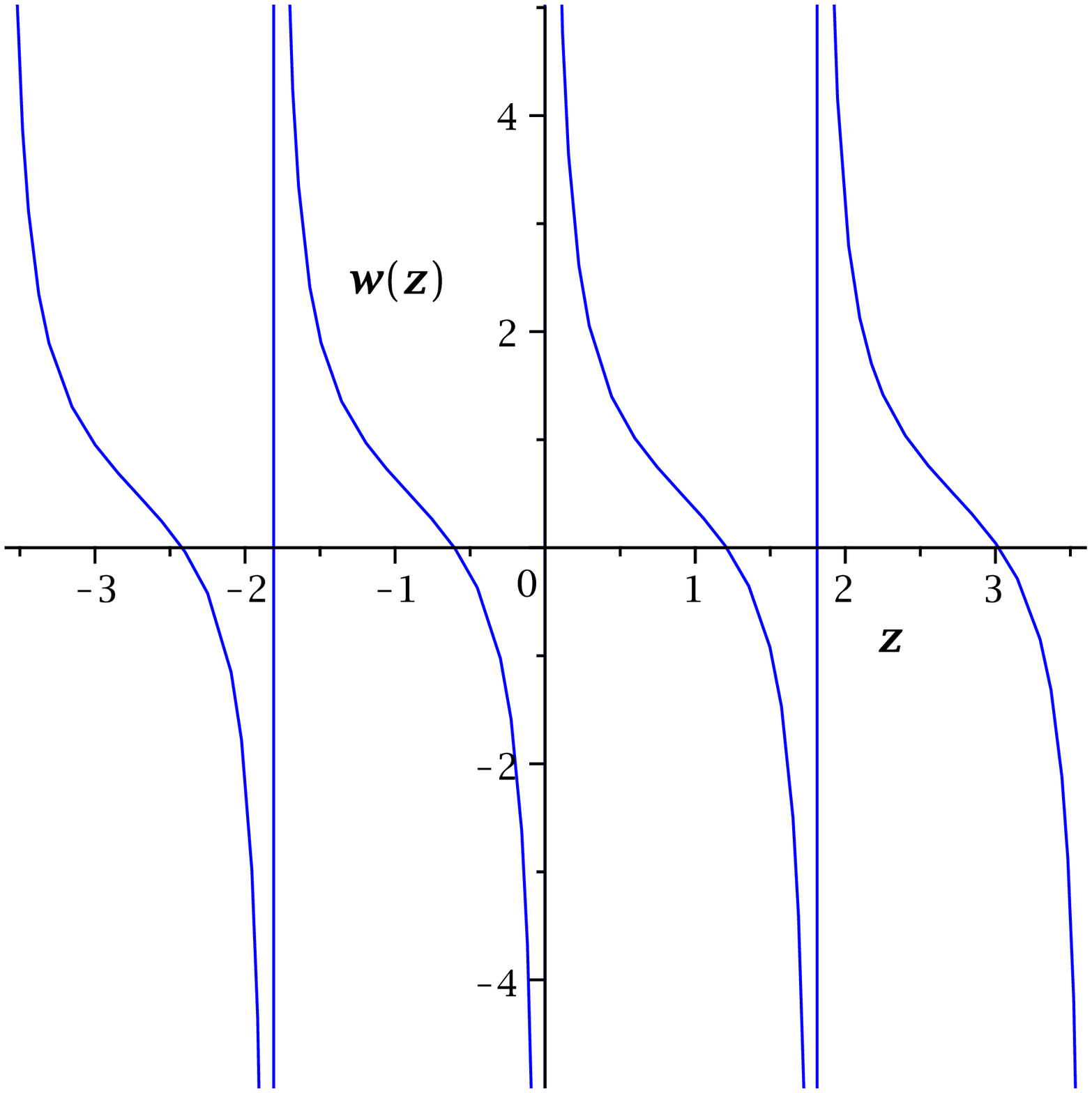,width=75mm}\label{Poles1_1}}
 \subfigure[$\alpha=10$, $\beta=10$, $z_0=0$, $T=\sqrt{93} \pi i/31$]{\epsfig{file=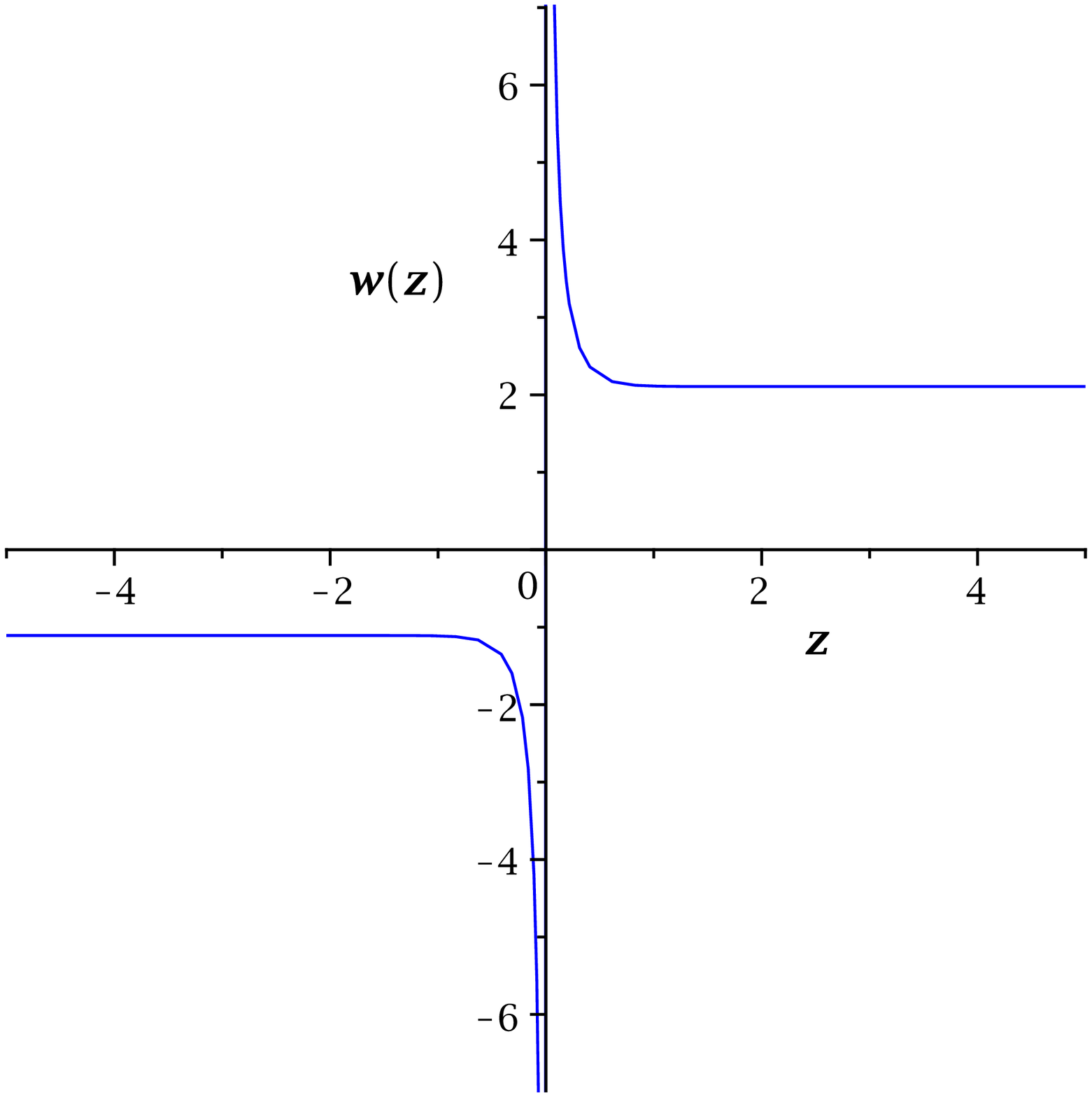,width=80mm}\label{Poles1_2}}}
 \caption{Family of solutions \eqref{Meromorphic_sols_example_Sol1}.}
 \label{F:Pole1}
\end{figure}
provided that the following correlation holds
\begin{equation}
\begin{gathered}
\label{Meromorphic_sols_example_gamma1}
\gamma=-\frac{(\beta^2+25\alpha)\beta}{375}.
\end{gathered}
\end{equation}
For the Laurent series \eqref{Meromorphic_sols_example_Exp2} we get
the family of meromorphic solutions
\begin{equation}
\begin{gathered}
\label{Meromorphic_sols_example_Sol2} w(z)=-\frac{2\pi}{3T}\cot
\left\{\frac{\pi(z-z_0)}{T}\right\}+\frac{2\beta}{33}, \quad
T=22\left(-\frac{2}{20\beta^2+363\alpha}\right)^{1/2}\pi
\end{gathered}
\end{equation}
and conditions on the parameters of equation
\eqref{Meromorphic_sols_example_equation}
\begin{equation}
\begin{gathered}
\label{Meromorphic_sols_example_gamma2}
\gamma=-\frac{(4\beta^2+121\alpha)\beta}{2662}.
\end{gathered}
\end{equation}
Note that in expressions \eqref{Meromorphic_sols_example_Sol1},
\eqref{Meromorphic_sols_example_Sol2} the period $T$ depends on the
parameters $\alpha$ and $\beta$ and can be real or complex. For the
functions with purely imaginary period $T=i\tau$, $\tau\in
\mathbb{R}$ the following correlation holds
\begin{equation}
\begin{gathered}
\label{Meromorphic_sols_formula1} \frac{\pi}{T}\cot\left(\frac{\pi
z}{T}\right)=\frac{\pi}{\tau}\coth\left(\frac{\pi z}{\tau}\right).
\end{gathered}
\end{equation}

\begin{figure}[t]
 \centerline{
 \subfigure[$\alpha=-10$, $\beta=10$, $z_0=0$, $T=22\pi/ \sqrt{815}$]{\epsfig{file=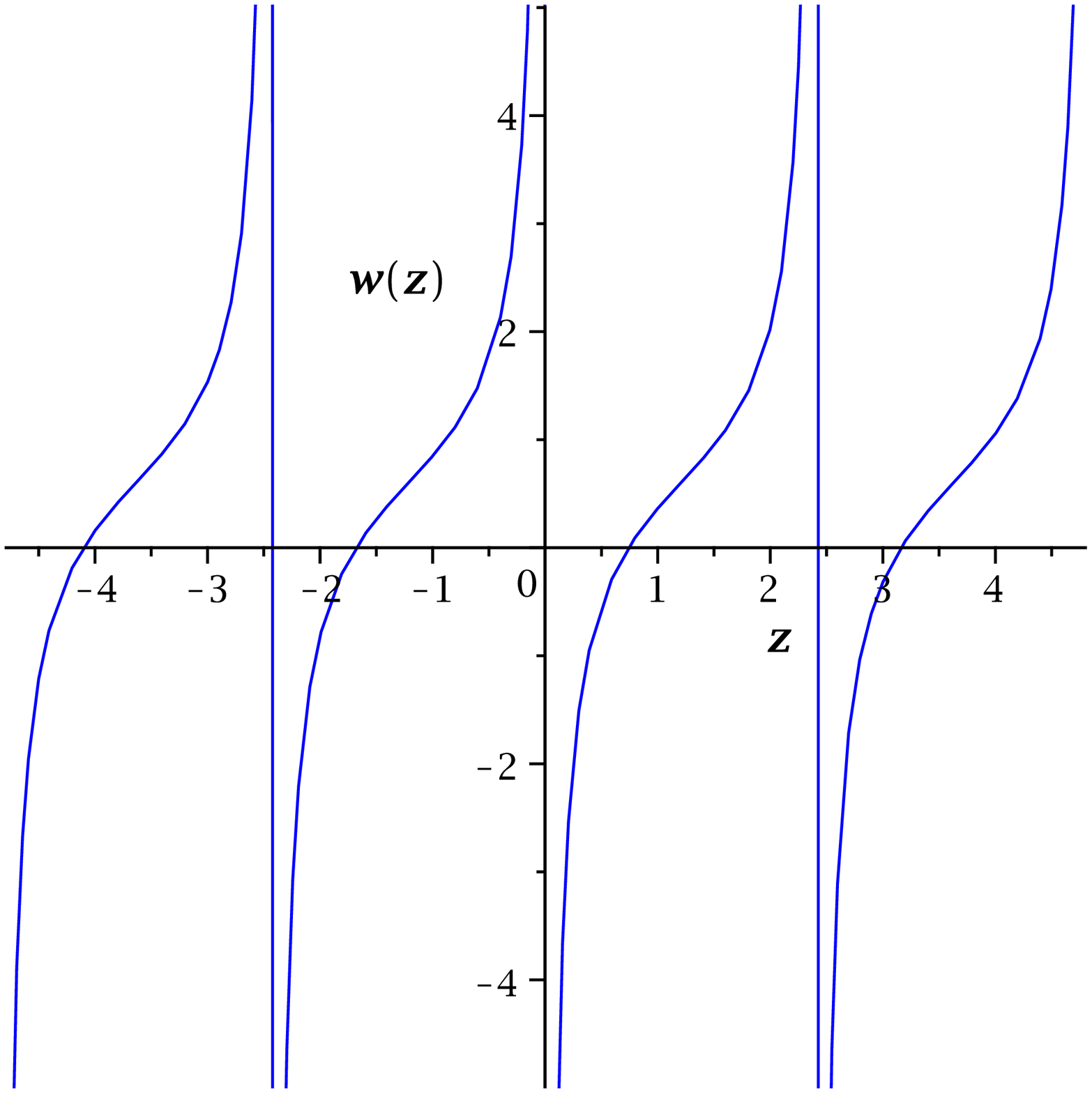,width=75mm}\label{Poles2_2}}
 \subfigure[$\alpha=10$, $\beta=10$, $z_0=0$, $T=22\pi i/ \sqrt{2815}$]{\epsfig{file=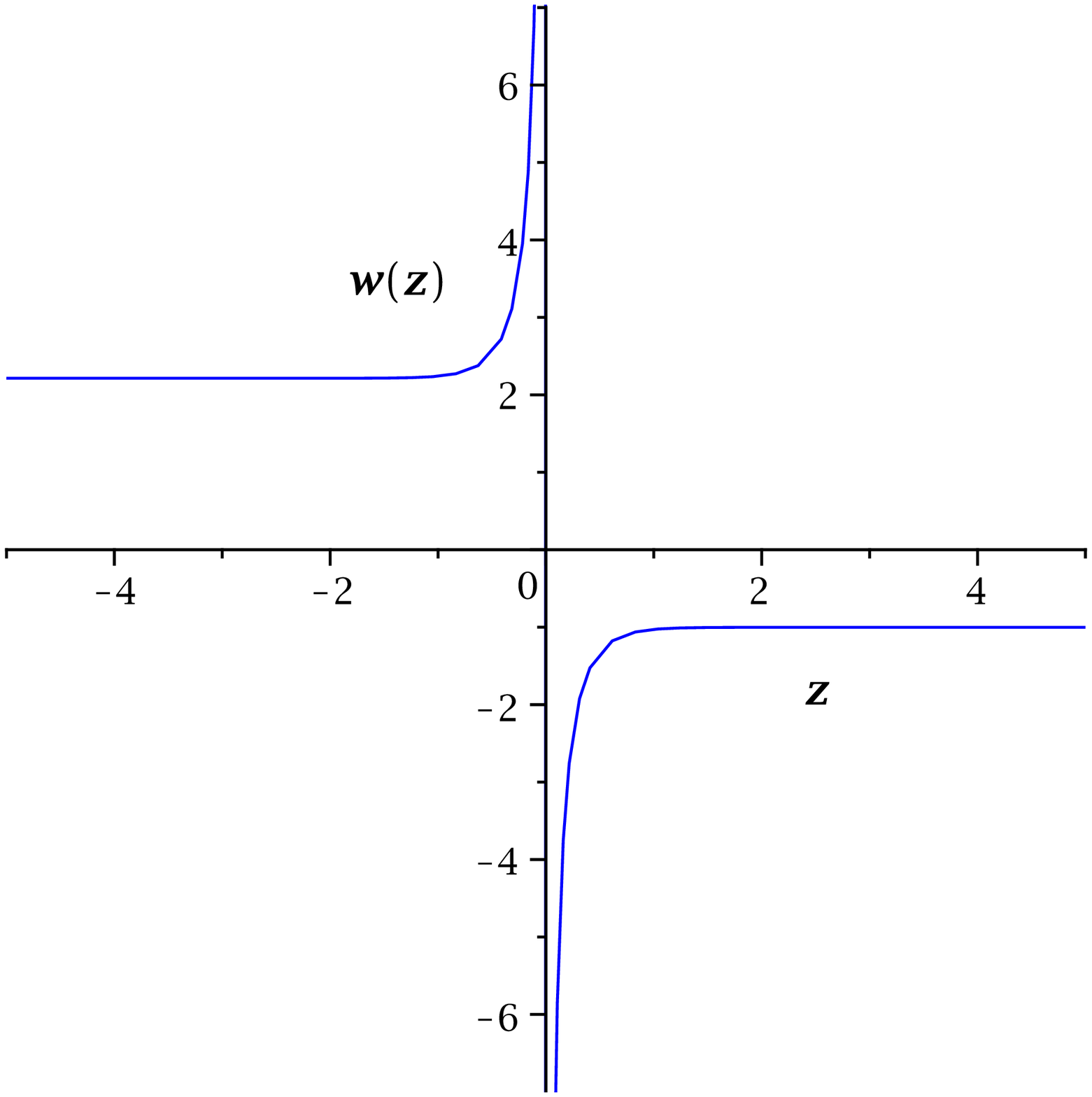,width=80mm}\label{Poles2_2}}}
 \caption{Family of solutions \eqref{Meromorphic_sols_example_Sol2}.}
 \label{F:Pole2}
\end{figure}

Now let us construct exact meromorphic solutions with poles of
different types. The general expression for this solution is the
following (see expression \eqref{Ex_Sol_ExppN} with $i_0=1$,
$I=\{2\}$, $a_2=a$)
\begin{equation}
\begin{gathered}
\label{Meromorphic_sols3} w(z)= c_{-1}^{(1)}\sqrt{L}\cot
\left\{\sqrt{L}z\right\}+ c_{-1}^{(2)}\sqrt{L}\cot \left\{\sqrt{L}
(z-a)\right\}+h_0.
\end{gathered}
\end{equation}
Again we omit arbitrary constant $z_0$. Without loss of generality,
the point $z=0$ is a pole of the first type, while the point $z=a$
is a pole of the second type. Constructing the Laurent series for
solution \eqref{Meromorphic_sols3} in a neighborhood of the points
$z=0$, $z=a$, we obtain
\begin{equation}
\begin{gathered}
\label{Meromorphic_sols_example_Exp4} w(z)=
\frac{c_{-1}^{(1)}}{z}+h_0-Ac_{-1}^{(2)}-
\left(c_{-1}^{(2)}(L+A)+\frac{c_{-1}^{(1)}L}{3}\right)z+o(|z|),\hfill
\\
w(z)= \frac{c_{-1}^{(2)}}{z-a}+h_0+Ac_{-1}^{(1)}-
\left(c_{-1}^{(1)}(L+A)+\frac{c_{-1}^{(2)}L}{3}\right)(z-a)+o(|z-a|),
\end{gathered}
\end{equation}
where we set $A\stackrel{def}{=}\sqrt{L}\cot\sqrt{L}a$. Comparing
the first of these series with the expansion
\eqref{Meromorphic_sols_example_Exp1} and the second
--- with expansion \eqref{Meromorphic_sols_example_Exp2}, we find
the parameters $h_0$, $T$, $A$ and conditions on the parameters
$\alpha$, $\beta$, $\gamma$ of equation
\eqref{Meromorphic_sols_example_equation}. Thus we have
\begin{equation}
\begin{gathered}
\label{Meromorphic_sols3_exact_new} w(z)=\frac{\beta}{220}\cot
\left\{\frac{\beta(z-z_0)}{110}\right\} -\frac{\beta\left(7\cot
\left\{\frac{\beta(z-z_0)}{110}\right\}+1\right)}{165\left(\cot
\left\{\frac{\beta(z-z_0)}{110}\right\}+7\right)}+\frac{61\beta}{660}.
\end{gathered}
\end{equation}
The family of meromorphic functions
\eqref{Meromorphic_sols3_exact_new} solves equation
\eqref{Meromorphic_sols_example_equation} provided that
\begin{equation}
\begin{gathered}
\label{Meromorphic_sols3_exact_conditios}
\alpha=-\frac{94\beta^2}{3025},\quad
\gamma=-\frac{31\beta^3}{33275}.
\end{gathered}
\end{equation}
Expression \eqref{Meromorphic_sols3_exact_new} can be rewritten as
\begin{equation}
\begin{gathered}
\label{Meromorphic_sols3_exact} w(z)=\frac{\beta}{220}\cot
\left\{\frac{\beta(z-z_0)}{110}\right\}- \frac{\beta}{165}\cot
\left\{\frac{\beta(z-z_0)}{110}+\frac12\log
\frac43\right\}+\frac{61\beta}{660}.
\end{gathered}
\end{equation}
Meromorphic solutions \eqref{Meromorphic_sols3_exact_new} with
$z_0=0$ possess two poles on real axes, one -- at the origin,
another one -- at the point $a=-55\log(4/3)/\beta$, $\beta\in
\mathbb{R}$. All other poles lie in the complex domain since the
period of solutions is $T=110\pi i/\beta$.

Now let us construct rational solutions of the equation in question.
Substituting $w=z^m$ as $z$ tens to infinity into equation
\eqref{Meromorphic_sols_example_equation} we see that $m=0$. Note
that at certain conditions on the parameters $\alpha$, $\beta$,
$\gamma$ expansions \eqref{Meromorphic_sols_example_Exp1},
\eqref{Meromorphic_sols_example_Exp2} terminate. As a result we find
two families of rational solutions and correlations for the
parameters of equation \eqref{Meromorphic_sols_example_equation}
\begin{equation}
\begin{gathered}
\label{Meromorphic_sols3_exact_rational1}
w(z)=\frac{1}{2(z-z_0)}+\frac{\beta}{20},\quad
\alpha=-\frac{11\beta^2}{200},\quad \gamma=\frac{\beta^3}{1000}\hfill\\
w(z)=-\frac{2}{3(z-z_0)}+\frac{2\beta}{33},\quad
\alpha=-\frac{20\beta^2}{363},\quad \gamma=\frac{4\beta^3}{3993}.
\end{gathered}
\end{equation}
If we try to find rational solutions with poles of two types, we
obtain an inconsistent system of algebraic equations. Consequently,
rational solutions of the form
\begin{equation}
\begin{gathered}
\label{Rat_sol_two_types}
w(z)=\frac{c_{-1}^{(1)}}{z-z_0}+\frac{c_{-1}^{(2)}}{z-a-z_0}+h_0.
\end{gathered}
\end{equation}
do not satisfy equation \eqref{Meromorphic_sols_example_equation}.

\begin{figure}[t]
 \centerline{
 \subfigure[$\beta=10$, $z_0=0$, $T=11\pi i$]{\epsfig{file=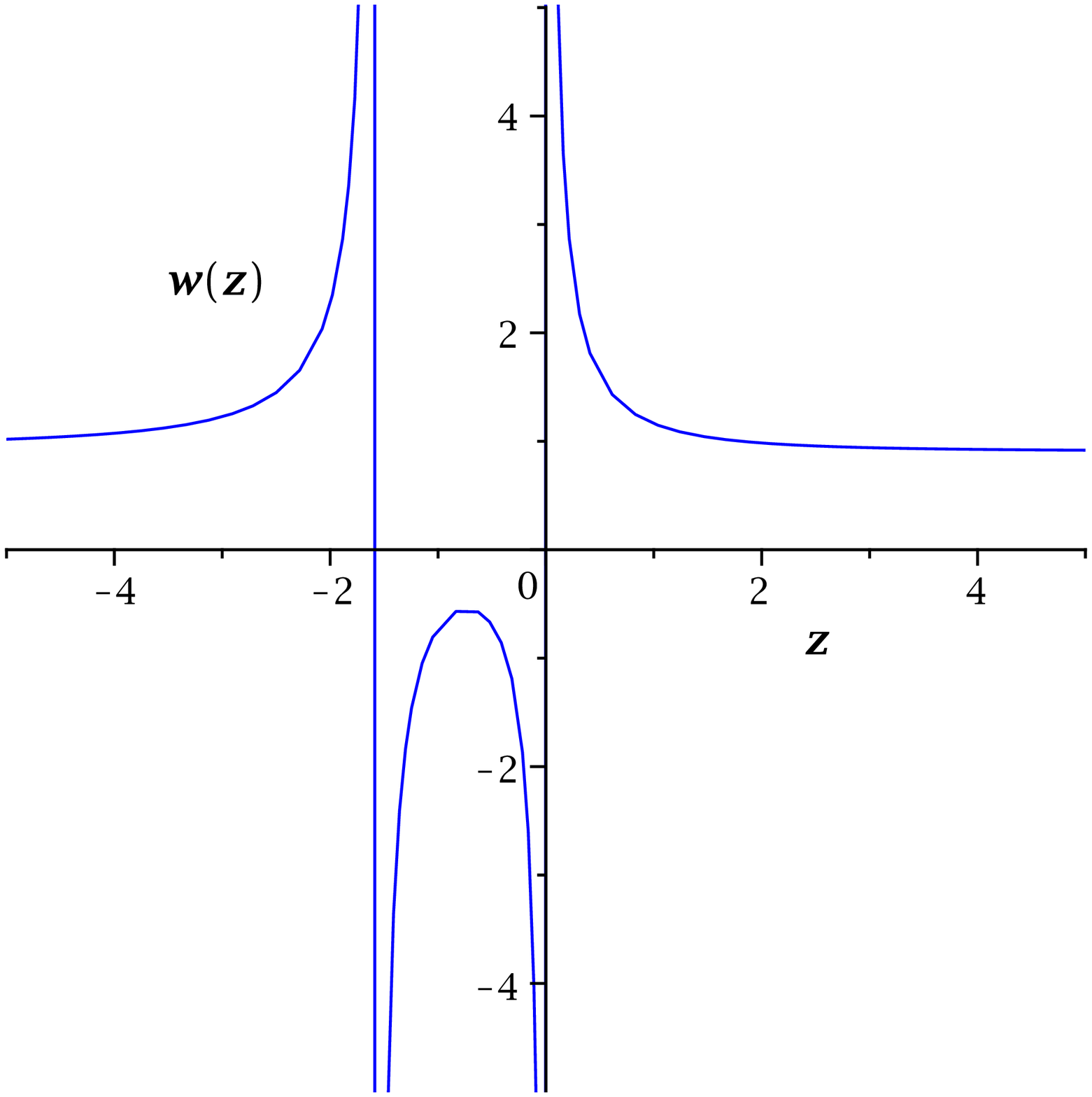,width=75mm}\label{Poles3_1}}
 \subfigure[$\beta=-10$, $z_0=0$, $T=11\pi
 i$]{\epsfig{file=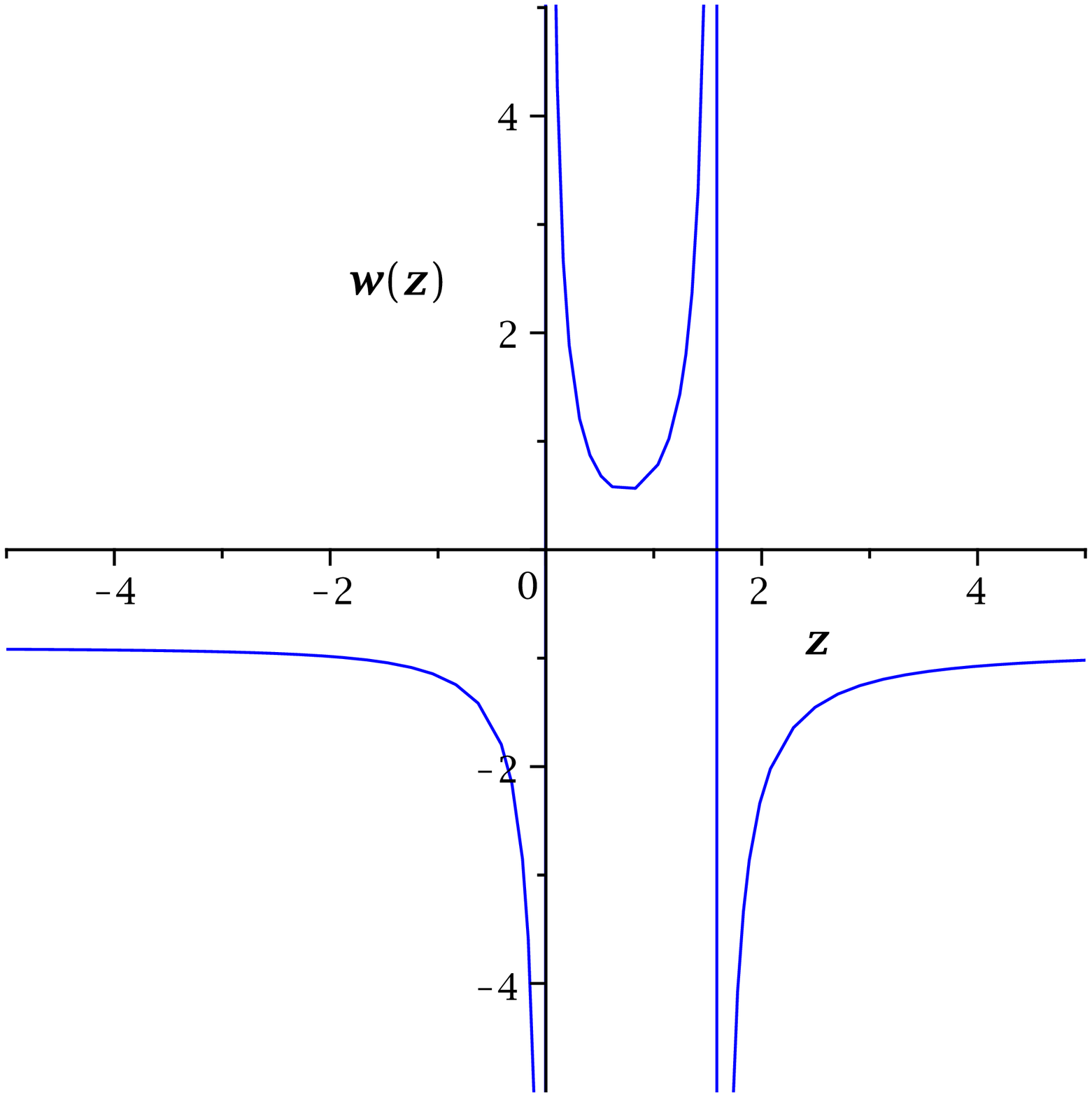,width=75mm}\label{Poles3_2}}}
 \caption{Family of solutions \eqref{Meromorphic_sols3_exact_new}.}
 \label{F:Pole3}
\end{figure}

Equation \eqref{Meromorphic_sols_example_equation} does not posses
elliptic solutions, since necessary condition
\eqref{Condition_ellipticN} does not hold.

We have studied all the expressions for meromorphic solutions given
in theorem 2. Hence, we have found all families of
nonconstant meromorphic solutions of equation
\eqref{Meromorphic_sols_example_equation}. Plots of these solutions
for certain values of the parameters are given in figures
\ref{F:Pole1} --- \ref{F:Pole3}.

\section {Conclusion}

In this paper we have presented a method for constructing exact
solutions of autonomous nonlinear differential equations. Our
approach is simple in application since one can introduce parameters
of exact solutions in such a way that for many equations linear
correlations on the parameters arise at each step. Our method
generalizes several other methods (for example, exp--function
method, tanh-function method \cite{Malfliet01} and their extensions
and modifications \cite{Parkes01}). A great drawback of these
methods is an a priori assumed expression for an exact solution. As
a consequences solutions outside stated family may be lost. Our
approach takes into account the structure of singularities and
possible Laurent expansions for exact solutions. Along with this, we
have obtained the general form of meromorphic solutions for a wide
class of autonomous nonlinear ordinary differential equations.

\section {Acknowledgements}

This research was partially supported by Federal Target Programm
"Research and Scientific - Pedagogical Personnel of innovation in Russian Federation on 2009 -- 2013 (Contract P2457).

\end{document}